# Implementing an ERP System: The Paradigm of a Chemical Company

Emmanouil Kolezakis

The Greek Ministry of Education

ekolezakis@sch.gr

**Abstract**: An application of a methodology present in the previous paper ["An ERP Implementation Method: Studying a Pharmaceutical Company" arXiv:1901.01810v1 [cs.SE]], is under consideration for a chemical company. Half of the paper as far as the methodology used is similar and only the application is different. In this paper it is tried to resolve problems and drawbacks of the methodology and to prove that it is satisfied enough to add some knowledge in the area of ERP systems Implementation. The modelling tool is again Petri-nets and the effort is to analyze the problem using this mathematical element. Finally, the SAP implementation method is presented, representing all the knowledge captured from this case study.

## 1. Introduction

ERP systems provide a software solution comprising several interconnected modules covering most of the key functions. For example SAP has modules for human-resource, material logistics, treasury, etc.This paper proposes a framework for the treatment of goal acquisition, alignment and reuse within the enterprise in order to facilitate the use of SAP. In the following section the ontology of the Reusable Organisational Change (ROC) framework is presented through the case of Electro Tech.

"Electro Tech is a fictional company, created from the collection of a variety of true life business practices" [Hiquet 1998]. It is manufacturer of electrical components and factory automation products. It has been in the business since late 50s and has demonstrated a consistent growth during 60s, 70s and 80s. The problems it faces are not unique but typical of a company who come across, IT evolution, globalisation, integration, mergers etc.

## 2. Ontology of the Methodology

The ROC framework consists of four static affinities namely:
(1) Organisational Goals
(2) Business Process Models
(3) Project Deliverables
(4) Requirement Reuse Plan.

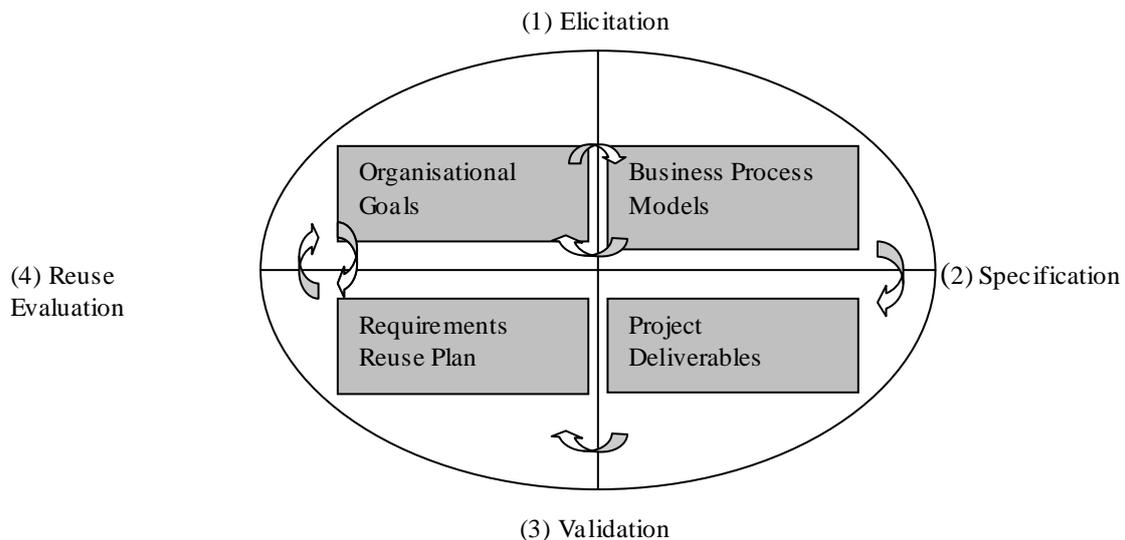

Figure 1: ROC Application Process

It consists of four dynamic affinities as well:
(1) Elicitation
(2) Specification
(3) Validation
(4) Reuse Evaluation

The elicitation phase of the framework includes the determination of the high level objectives of the enterprise as well as the more functional one. Product of this phase is the goal-graph notation (see fig.3).
In the second stage the outcome is the Petri-net notation of the As-Is state. In the third stage we can see the alignment of the processes of both SAP and enterprise. In this stage we have a more concrete idea of what to expect from the project itself. Product of this stage is the Petri-net notation (The To-Be state).
Finally the Reuse Evaluation phase stores the knowledge captured during the project implementation process for future use. Especially stores the strategy followed during the third phase of the framework so that we use this knowledge in possible similar forthcoming projects.

## 2.1 Goal Elicitation Sub-model

The goal-elicitation sub-model is illustrated in fig.2. Central to this view is the concept of organisational goal. An organisational goal is a desired state of affairs that needs to be attained. Typical goals in the case of Electro Tech are: "*improve IS services*" or "*Automate payroll*" or "*satisfy customer need for information from their suppliers*" etc.

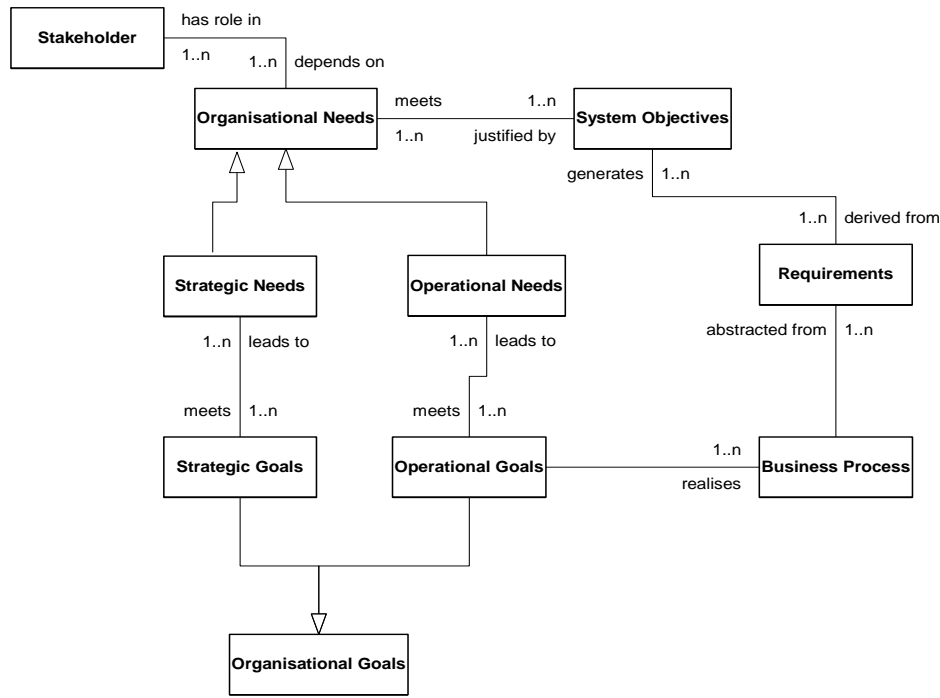

Figure 2: Goal Elicitation Sub-model

Goals are pertain to stakeholders. A stakeholder is defined as someone who has an interest in the system design and usage. Examples of stakeholders are: managers, system designers, system users, customers etc.

Systems are built to primarily satisfy organisational needs. These needs may either be long term strategic needs or more immediate operational needs. These two needs support each other and are often more detailed in a goal hierarchy. We must distinguish between goals and needs. Goals derived from needs which are expressed in a more abstract manner. For example "*need for information*" is a general organisational need for Electro Tech. From this need derived the goal "*automate payroll*" and "*satisfy customer need for information from suppliers*"

System objectives elicited from organisational needs and are determined by stakeholders. For example the objective: "*supply with the latest technology*" is an objective of the system determined by stakeholders, probably management but depends on organisational needs for leading edge technology. Operational needs lead to the formulation of operational goals and strategic needs to the formulation of strategic goals. For example "*Buy a VP of sales and marketing state of the art system*" is a strategic goal whereas "*Buy a VP of sales and marketing system using lotus 1-2-3*" is an operational one.

The requirements for the system are generated from system objectives because high level goals are too vague to be called requirements. Requirements must be more specific to proceed further.

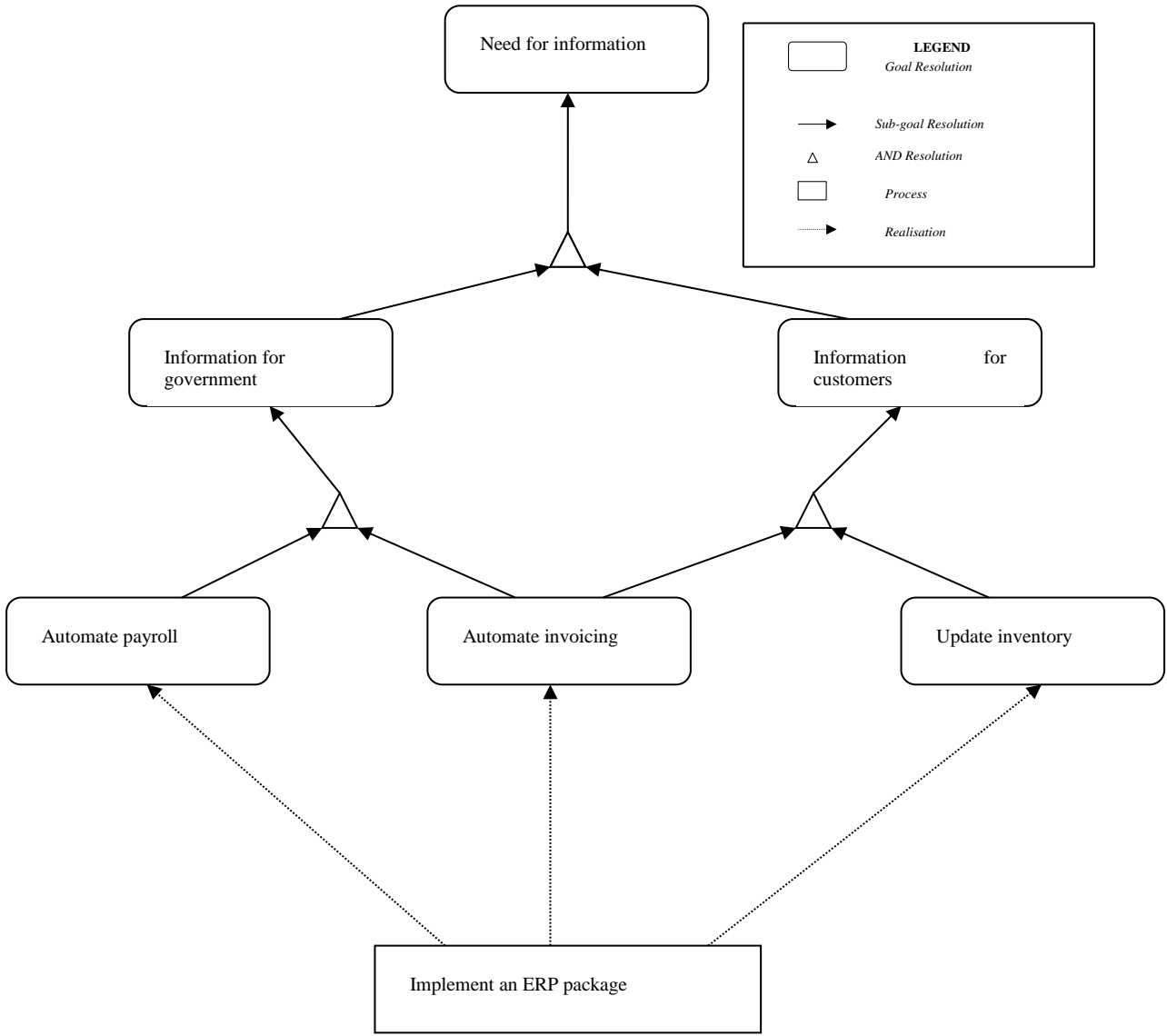

Figure 3: Goal-Graph for Electro-Tech

In fig. 3 we have an example of the elicitation of the enterprise goals in a notation as it has been used from [Loucopoulos et Kavakli 1997]. We start from the high level objectives of the enterprise such as "*need for information*" and we end with the low level objectives, more concrete goals such as "*autmate payroll*", "*automate invoicing*" and "*update inventory*". These goals could lead as it has been stated before into operational goals such as "*Buy a VP of sales and marketing system using lotus 1-2-3*". This goal graph depicts mainly the strategic goals derived from strategic needs. The realisation of these goals comes from the process "implement an ERP package" in our case SAP. If we proceed further from the strategic goals we have the elicitation of the operational goals. The operational goals are realised by the business.

## 2.2 Specification of the Business Processes Sub-model

The specification sub-model is illustrated in fig.4. Central to this meta-model are the change goals. Change goals are elicited from current organisational goals and with the help of Business Process Models.

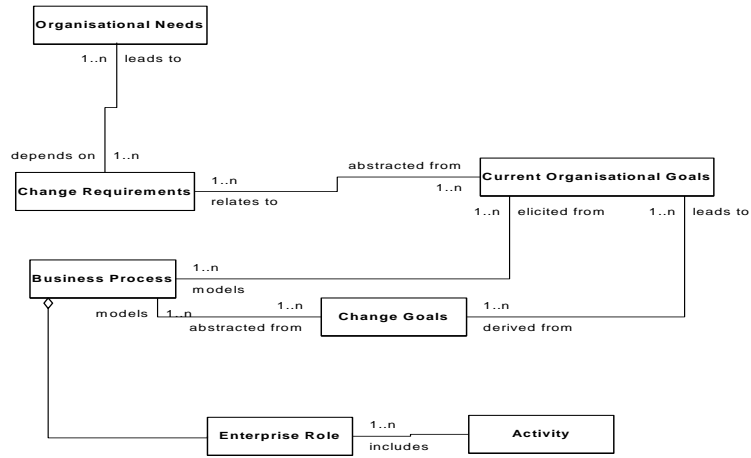

Figure 4: Specification Sub-model.

.

Organisational needs leads to change requirements. For example the "*need for an integrated IS because of very diverse citation*" leads to the change requirement "*improve MIS services*". That means that we model the business process then the goals are derived from the models and are specified in Petri-nets notation. A current organisational goal can be "*satisfy customer need for information from suppliers*" and a change goal "*develop a homegrown IS*".

## 2.3 Validation Sub-model

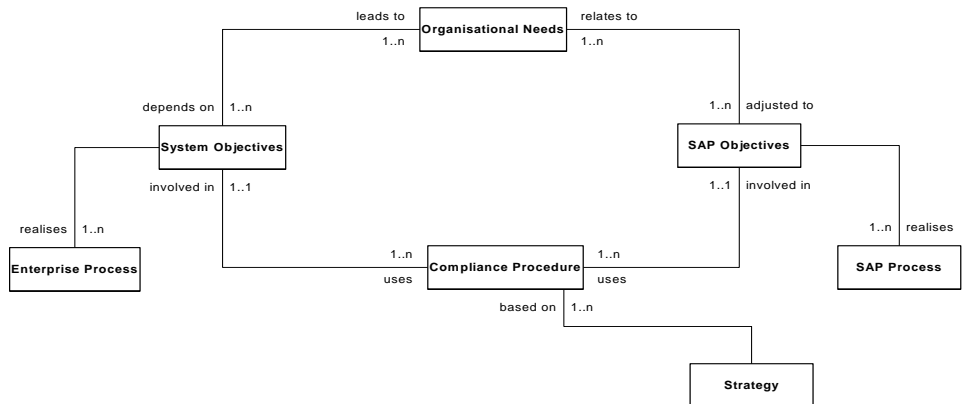

Figure 5: Validation Sub-model

In this stage it is being examined if the business objectives can be fulfilled and what changes must be made in the business processes of the enterprise in order to realise these. This is an interactive process that transforms the enterprise's business requirements into a future SAP solution. This interactive process provides continuous feedback mechanisms that initially identify gaps and then evolve to filling the gaps.

Installations of ERP systems are difficult to align to specific requirements of the enterprise because of the low level at which functionality is described. Central to this metamodel is the compliance procedure. The following terms are defined:

SAP goals: The tasks carried out by a SAP function.
SAP process: The process who realises a SAP goal.
SAP strategy: The combination of all necessary SAP processes, in order to reach a SAP goal.

The main reason for thinking in terms of goals (intentional level) and strategies (Strategy level) is that we need a common way of communication between SAP and enterprise. Organisations think in terms of their objectives and their strategies and SAP functions have a supportive role. SAP goals must support and implement enterprise goals.

## 2.4 Reuse Evaluation

The purpose is to reduce significantly the task of creating application-specific models and systems: the user select relevant parts of the reference model, adapts them to the problem at hand, and configures an overall solution from these adapted parts. Since the analysis of a domain can take an enormous effort when started from scratch, the use of reference models has been reported to save up to 80% in development costs for systems in standardised domains [Scheer 1994]. According to Ramesh and Jarke [Ramesh and Jarke 2001] reference models have become highly successful in many industries and among the best known examples is the SAP approach. Each case can be consider a scenario S. The retrieval of a scenario S can follow the algorithm [Aamodt and Plaza 1994]:

A. **Problem create New Case**
B. **Retrieve an old similar Case**
C. **Compare Retrieved Case and New Case**
D. **Reuse Solved Case**
E. **Test Suggested Solution**
F. **Retain Learned Case**
 Next

## 3. Aligning ERP to Enterprise

In this section we examine and model in more detail the stage between (1) elicitation and (3) validation of the framework whereas the alignment of both SAP and enterprise processes takes place and the most vital work is being done. We study the current state of the enterprise (As-Is) and then we study the desired state (To-Be) of the proposed SAP solution. Both states are representing in Petri-nets, a modelling element where the correspondent strategy for each transition is the key feature for the alignment.

## 3.1 Logical Organisation

The levels of abstraction are (Figure 6):

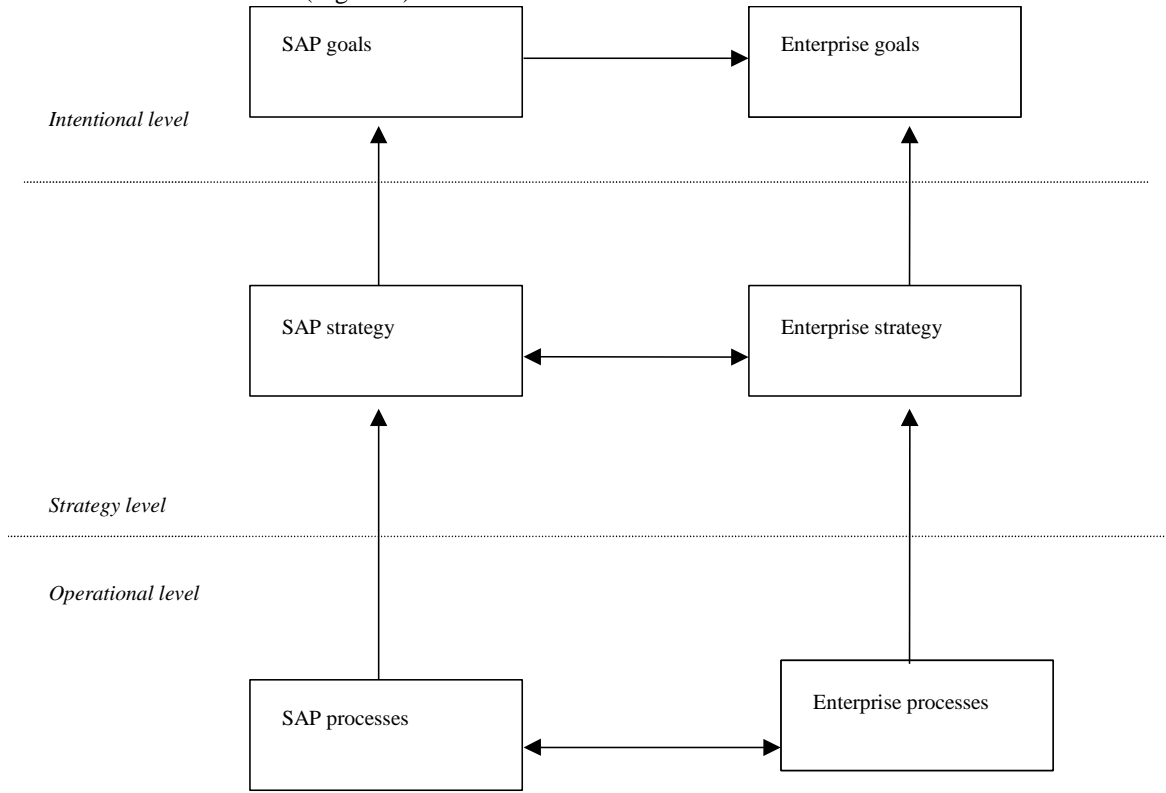

Figure 6: Levels of Abstraction

(a) Intentional level
(b) Strategy level
(c) Operational level

The Intentional level defines concepts such as goals. The SAP goals must be supportive to the enterprise goals and their realization must help towards the realization of the enterprise objectives.

The Strategy level defines concepts such as SAP strategy or enterprise strategy.

The Operational level includes concepts such as SAP processes or enterprise process, namely the process that realizes the enterprise goals.

So we have alignment of the processes in a strategy level, expressed in Petri-nets and the SAP goals support enterprise's high level objectives.

## 3.2 Modelling the Current State

Petri-nets [Petri 1962] are to model the current state of the enterprise, in our case Electro Tech. The production planning module (PP) of SAP is used as an example. The PP module is a flexible module containing several alternative strategies adjusted to each particular enterprise according to its objective and the targeted process the stakeholders want to implement. Because of the modularity which distinguishes SAP the Petri-net notation is a particularly suitable modeling tool for ERP systems in our case SAP.

Each node in the Petri-net notation corresponds to a state in the production planning process. Using the triplet form <source state, target state, strategy> we can have the correspondent textual notation for the transmission from a place to another place. This graphical notation used to represent the correspondent fragments of the SAP production planning process, having same milestones but probably different strategy can be used for adjusting the SAP planning strategy into the targeted enterprise process. The current status of the enterprise as it is described is [Hiquet 1998]:

(a) There is a serious problem in the control of approve vendors which is fragmented and controlled manually. The manual purchasing system can cause errors at the first place and at the second, there is no history option or any possibility for anticipation. As a result the Sales and Operations plan is created manually.
(b) There is no demand management strategy in the supply of raw material. That means no forecasting strategy within the supply chain.
(c) There is no real time production planning strategy. That means need for production in advance and application of a stock strategy. (Make-to-Stock strategy).
(d) Lack of an on-line order processing system even automated.

As it comes up from the previous conclusions the identified problems are fall into two categories:

- Supply Chain Management problems, for example (a) and (b),
- Demand Side problems, for example © and (d).

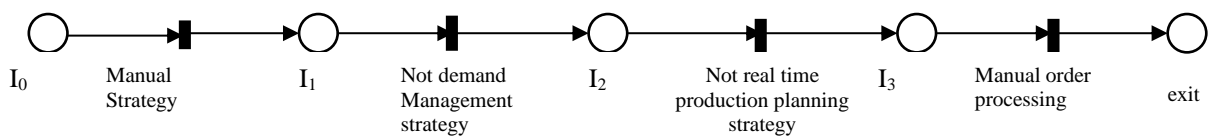

Figure 7: The Current Status of Electro Tech.

$I_0$: start, $I_1$: support material, $I_2$: work with material, $I_3$: Stock.

Using the notation described previously we can depict all the above making thus possible the comparison or better the adjustment of the future SAP solution into business processes of the enterprise (fig.7)

Each fragment of the above representation corresponds to a business situation, illustrating the problems stated. We use the triplet form <source state, target state, strategy> to represent the transition from one place to another place, and the strategy element depicts how this is being done.

In the current status of Electro Tech the fragments are:

**PF$_1$** :<(start), (support material), *manual strategy*>,
**PF$_2$** :<(support material), (work with material), *Not demand management strategy*>,
**PF$_3$** :<(work with material), (stock), *Not real time production planning strategy*>,
**PF$_4$** :<(Stock), exit, *manual order processing strategy*>.

In a few words we can describe the functionality of the company as:

**PF$_1$** :<(start), (support material), *manual strategy*>: Production scheduling is reacting to the occurrences in the plant. All planning strategy is based on word-of-mouth.

**PF$_2$** :< (support material), (work with material), *not demand management strategy*>:
As commonly defined, demand management is the function of recognising all demands for a product, including forecasts, customer orders, interplant orders, etc. The output of demand management is referred to as the demand program. It consists of a list of independent requirements for each material specifying the quantities and dates the material is needed. Multiple versions of the planned independent requirements can be useful for segregating and managing different components of demand. In Electro Tech the sales demand is composed of direct sales and warranty replacement. The warranty replacement demand is calculated manually based on reliability data and total number of units in service.

**PF$_3$** :<(work with material ), (stock), *not real time production planning strategy*>: As a result of the previous problems there are delays in the production process of Electro Tech. Especially when a company has several plants depended one to another as far as the production process is concerned.

**PF$_4$** :<(stock), exit, *manual order processing strategy*>: What is really needed is a truly integrated system, without redundancy and with real-time transactional integration between business functions.

The correspondence between the identified problems and the process fragments is:

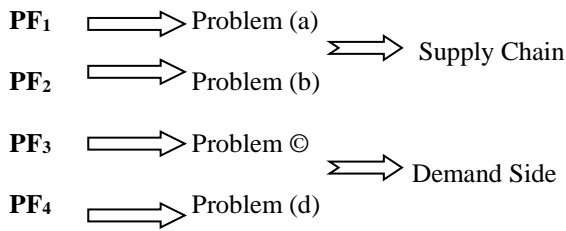

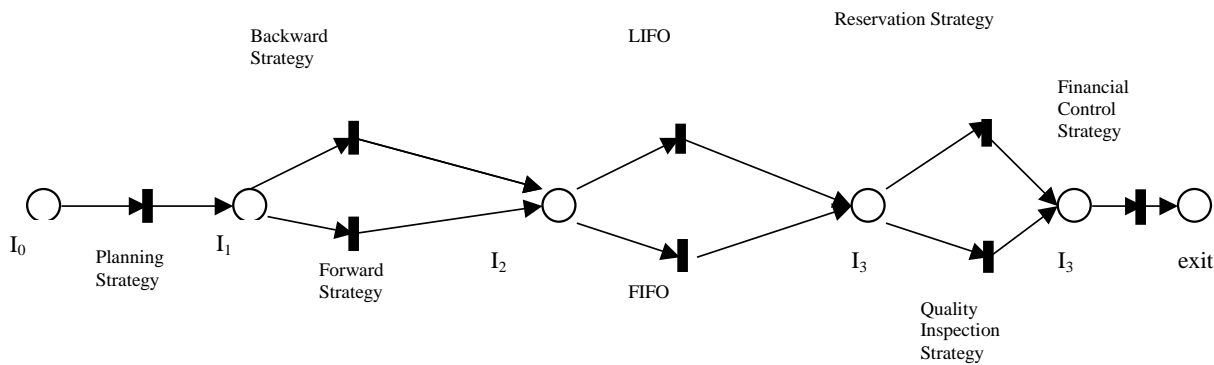

Figure 8: Proposed SAP Solution

## 3.3 Modelling the Proposed System

In the same manner we modeled the current state of the enterprise we model the future SAP solution. Because we are referred to Production Planning process and the Production Planning module of SAP is a highly flexible module consisting of several implementation strategies we must select an appropriate implementation strategy for the particular company. It is difficult to present a general overview of the SAP production planning process. The production planning functionality is a highly flexible collection of sub-modules and functionality that can be linked together to form a coherent planning and scheduling process. Every SAP implementation has the opportunity to use all or a part of the SAP planning functionality in order to meet organisations specific planning needs [Keller and Teufel 1998].

Figure 8 depicts an example of how the functional sub-modules or according to our determination components can be used in order the specific planning needs of the company to be fulfilled. In this example the sales representative create sales plan in the sales and operations component. These sales plans are then copied into Demand Management by the master production scheduler and smoothed out from weekly basis into daily basis. Once the master plan (in Demand Management) is satisfactory, master production scheduling is performed, followed by detailed material requirements planning.

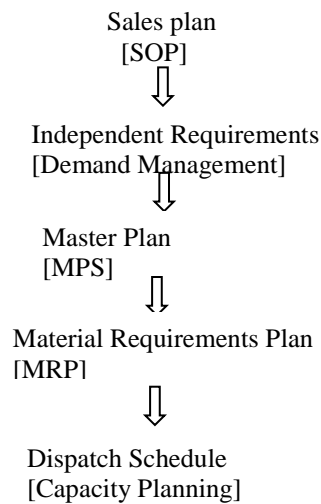

Figure 9: Sample SAP production planning process [Hiquet 1998].

The detailed material requirements are fed into the capacity planning system for finite production scheduling and dispatching.

This process would be applicable for a company that has accurate sales forecast information based on sales account representative feedback from customers, or some form of accurate market predictors. The production environment would typically be a job shop with relatively expensive finished goods being produced in a complex manufacturing and assembly process.

For Electro Tech we selected a make-to-stock strategy consisting of the Sales and Operations component, Master Processing Scheduling, Material Requirements Planning (MRP), Quality Management (QM), and Product Costing (PC) component. The representation in Petri-nets of the above module is the one in figure 8 whereas $I_0$:start, $I_1$:support material, $I_2$:work with material, $I_3$:Stock.

The business process in SAP concerning the PP module with the above planning strategy has the following fragments:

**PF₁**<(start), (Support material), *planning strategy*>
**PF₂**<(Support material), (work with material), *backward strategy*>
**PF₃**<(Support material), (Work with material), *forward strategy*>
**PF₄**<(Work with material), (stock product), *LIFO*>
**PF₅**<(Work with material), (stock product), *FIFO*>
**PF₆**<(Stock Product), (Stock Product), *Reservation Strategy*>
**PF₇**<(Stock Product), (Stock Product), *Quality Inspection Strategy*>
**PF₈**<(Stock Product), exit, *Financial Control Strategy*>.

Each of the above fragments treats a specific problem of the enterprise. For example the fragments PF1, PF2, PF3, PF4, PF5 resolve problems related to supply chain, whereas the fragments PF6, PF7, PF8 resolve problems related to demand side of the enterprise. The adjustment of the components of the Production Planning module of SAP has been done in such a manner so that to implement the selected planning strategy. Diagrammatically this can be proved as it is depicted below.

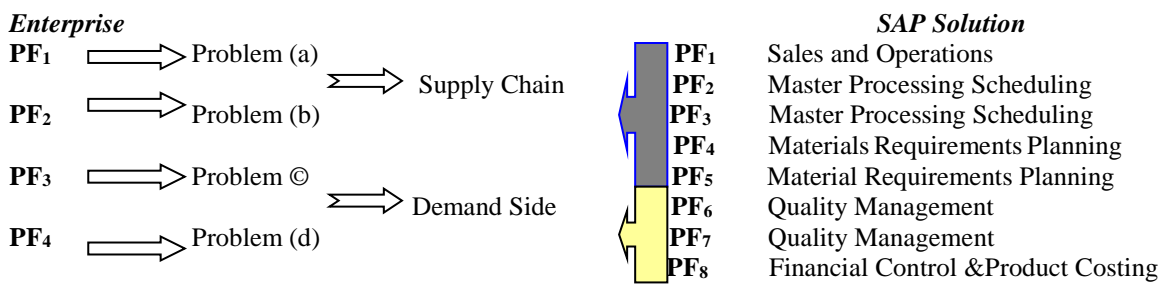

## 4. Application

We have seen so far the methodology and the ontology of the ROC framework for assisting mainly the alignment of the enterprise requirements to SAP requirements and the reuse of the captured knowledge. This section presents the empirical results and observations from applying the above framework in an industrial application capture from the literature.

The intention of this discussion is twofold: The first is to assess the applicability of the ROC framework on a non-trivial application. The second is to discuss a number of challenging issues that need to be addressed in order the proposed Framework to support real, complex tasks in the context of business modelling and reuse within the ERP domain and particularly SAP.

## 4.1 The Chemical Industry

This industrial application concentrates on a large-scale SAP R/3 implementation. ALVEO specializes in the marketing, sales, production and development of polyelefin foams and aims at being market leader in this field. Its organizational structure its decentralized but close knit.

## 4.1.1 Application Background

The company we study was founded in 1971 as a joint venture company with Swiss, Japanese and American partners [Welti 1999]. The headquarters for the Europe administration are located in Lucerne, Switzerland. The company has two production units, in the Netherlands and Wales and its sales distribution net, includes local sales offices in all key European markets.

In order to improve customer service and enforce sales, ALVEO has divided the market into three sectors:
- Industrial/Consumer sector (e.g consumer items, building and construction industry)
- Transportation sector (e.g carpet backing)
- Specialized sector (e.g Medical applications)

The company decided to replace its outdated computer system in all business areas with new software and appropriate hardware. The SAP implementation project was called FuturA standing for "The Future of ALVEO".

A totally new IT environment was desperately needed on the following grounds:
- Information need for management decisions
- Customer service complied with the increasing demands from customers
- Communication, both internal and external
- Software to accomplish company's growth.

Figure 22 depicts the acquisition of the enterprise goals. In this case the company decides to implement the whole SAP modules. This decision can be explained if we take into consideration the distributed environment into which the enterprise operates. Thus they need common procedures in order to succeed the desired integration between the different plants. The SAP goals are supportive to enterprise goals but the targeted process in this case has as elements SAP modules.

## 4.1.2 Current state Vs Proposed SAP Solution

The As-Is state of the enterprise as it is depicted from the problem it faces can be described as [Welti 1999]:
(a) Detailed information was needed as a ground for management decisions
(b) Better production cost transparency was necessary to improve the quality of decisions
(c) Shorter delivery times and reliable responses to increasing demand from customers was needed
(d) On-line access to stock and production data was required
(e) Demand for electronic communication with customers and suppliers.

The above problems are falling into the following categories which can be considered as the targeted processes:

(A) Logistics (a), (c), (d), (e)

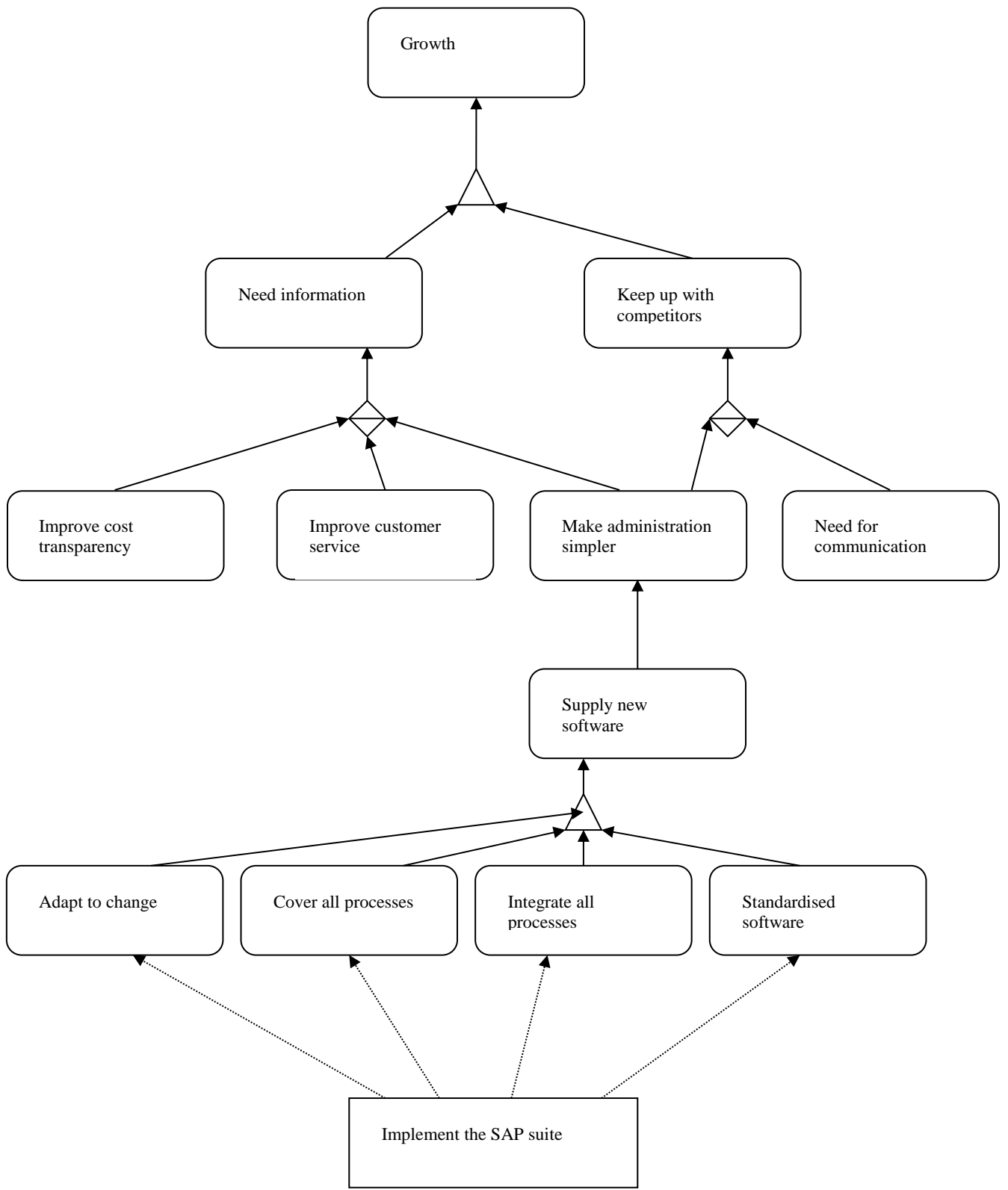

Figure 10: Acquisition of ALVEO's goals

(B) Accounting (b)

In order to resolve the above problems, the company decided to implement the following modules:
For the headquarters: Sales and Distribution (SD), Finance (FI) and Controlling (CO)
For the plants (Netherlands and UK): Production planning (PP), Material Management (MM), Finance (FI) and Controlling (CO).

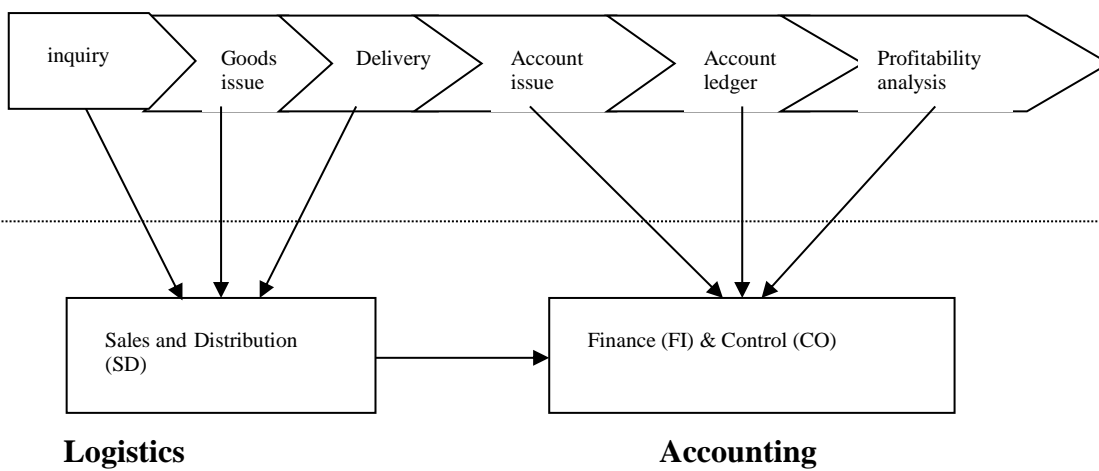

Figure11: Targeted process for the headquarters

The implementation strategy of the enterprise is a step-by-step strategy from one plant to another and a roll-out approach. A step-by-step implementation is characterized by the implementation of the software in small steps and normally concentrates on a few related modules at one time.

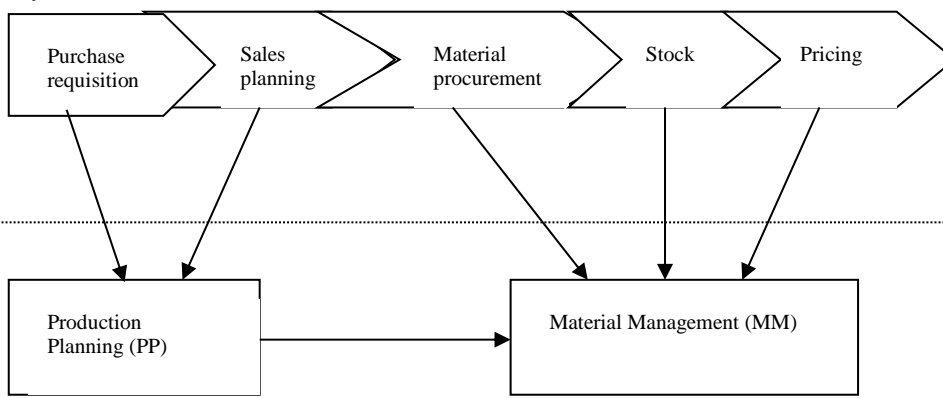

Figure12: Targeted logistics process for the plants

A roll-out approach creates a model implementation at one site which is then roll out to other sites [Welti 1999].

### 4.1.3 Headquarters

For the headquarters the logistics process includes the sales and distribution process and the accounting process includes financial accounting and controlling.

### A. Logistics

For the sales and distribution process the current state (As-Is) is :

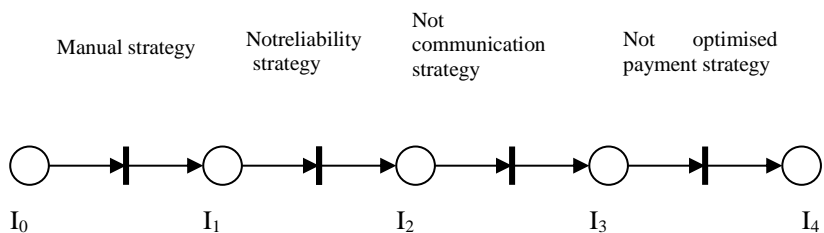

Figure 13: Current state for sales and distribution

$I_0$: Customer Inquiry, $I_1$: Quotation, $I_2$: Goods Issue, $I_3$: Goods Delivery, $I_4$: Billing.

In the above figure are illustrated all the problems about sales and distribution business process. For example there is a lack of communication, there is no reliability between the customers and the company, even within the company (e.g they can not check stock availability), and the payment method is not the best. The proposed SAP solution includes the installation of the complete Sales and Distribution (SD) module. The (SD) application module contains functions to manage activities such as selling products or performing services. Additionally it can perform the tasks of sales, shipping and billing [SAP 1994]. The Sales and Distribution module contains all of the functions important for sales, such as material determination, price determination, schedule determination, availability check, costing, credit limit checking, reservations, batch determination, order tracing, tax and condition determination, individual and collective invoice processing, returns processing, picking, transport planning and handling of foreign trade [Keller and Teufel 1998].
The future state with the (SD) module in the company will be:

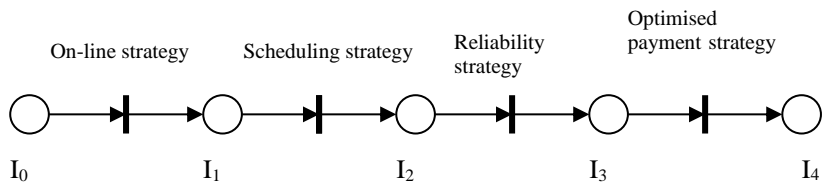

Figure 14: Future state for sales and distribution

$I_0$: Customer Inquiry, $I_1$: Quotation, $I_2$: Goods Issue, $I_3$: Goods Delivery, $I_4$: Billing.

The process fragments in each one of the above representations are:
Enterprise:
**$PF_1$**:<(customer inquiry), (quotation), (manual strategy)>
**$PF_2$**:<(quotation), (goods issue), (not reliability strategy)>
**$PF_3$**:<(goods issue), (delivery), (not communication strategy)>
**$PF_4$**:<(delivery), (billing), (not optimised payment strategy)>.

SAP:
**$PF_1$**:<(customer inquiry), (quotation), (on-line strategy)>
**$PF_2$**:<(quotation), (goods issue), (scheduling strategy)>
**$PF_3$**:<(goods issue), (delivery), (reliability strategy)>
**$PF_4$**:<(delivery), (billing), (optimised payment strategy)>.

The operational functional areas covered are:
- Operative sales support and shipping or transport. For example $PF_3$:<(goods issue), (delivery), (reliability strategy)>
- Request for quotation. For example $PF_1$:<(customer inquiry), (quotation), (on-line strategy)>
- Internet functionality and EDI. For example $PF_1$:<(customer inquiry), (quotation), (on-line strategy)>
- Credit limit checking. For example $PF_2$:<(quotation), (goods issue), (scheduling strategy)>
- Pricing and billing. For example $PF_4$:<(delivery), (billing), (optimised payment strategy)>.
- Contracts, scheduling agreements. For example $PF_2$:<(quotation), (goods issue), (scheduling strategy)>

The sub-modules which corresponds to the above fragments are:

| Process Fragments | Components |
|---|---|
| $PF_1$ | Sales |
| $PF_2$ | Shipping |
| $PF_3$ | Shipping |
| $PF_4$ | Billing |
| All | Sales and Distribution IS, Sales support |

Table 1: Correspondence between process fragments and components

The integration of the Sales and Distribution module in the R/3 system is also assured across boundaries, for example with Material Management (MM), with Production Planning and Control (PP), with Controlling (CO) and with Financial Accounting (FI). In order processing, an availability check with regard to the stocks in the Material Management module can be performed in the (SD) module. Delivery processing can create goods issues that are processed by the Material Management module. Likewise, in order processing, the existing sales quantities of one or more customer orders can be transferred to the Production Planning and Control module as planned independent requirements to be produced. In demand management, the existing customer orders are then included in production planning and, if necessary, consumed against the quantities from planning.

Furthermore, assembly orders for production can be created directly in the Sales and Distribution module. Another aspect of integration is the fact that, in order processing, the Sales and Distribution module can access the bills of materials from the Production Planning module and retrieve from corresponding values for price determination, availability checking and delivery date determination. If an enterprise's product consists of one piece products, the appropriate data from the Sales and Distribution module are passed on to Controlling for product costing purposes. During billing, the invoice amounts are automatically transferred, as revenues to the appropriate accounts in Financial Accounting [Weihrauch 1996].

## B. Accounting

The accounting process includes the financial accounting, and the controlling processes which are realised respectively, by the Financial Accounting (FI) module and the Controlling (CO) module. The above process is mainly the same for the headquarters and the plants and it is implemented during the first year of the project to all of them. The (As-Is) state of the enterprise for the accounting process is:

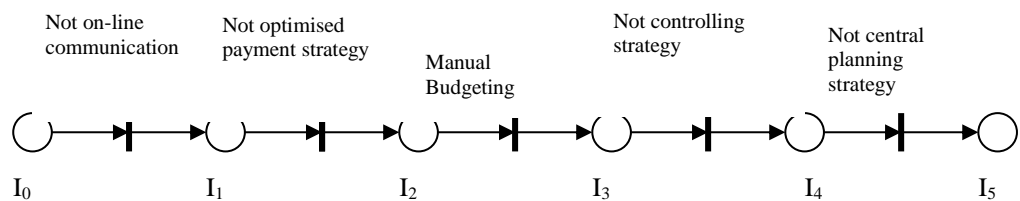

Figure 15: Current state for the accounting process

$I_0$: Order entry, $I_1$: Account issue, $I_2$: Accounts ledger, $I_3$: Process consumption accounting, $I_4$: Product cost accounting, $I_5$: Profitability analysis.

The problems the enterprise faces are mainly lack of communication and lack of automation for trivial business processes.
The future state with SAP's Financial Accounting (FI) and Controlling (CO) modules will be:

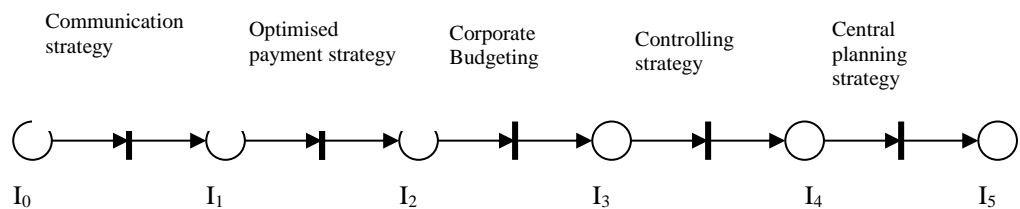

Figure 16: Future state with SAP for the accounting process

$I_0$: Order entry, $I_1$: Account issue, $I_2$: Accounts ledger, $I_3$: Process consumption accounting, $I_4$: Product cost accounting, $I_5$: Profitability analysis.

The main objective of the SAP implementation for the accounting process is to reduce payment terms by 5 days and to optimise routine administration work in finance.

The process fragments of the (To-Be) status:
$PF_1$: <(order entry), (account issue), (communication strategy)>,
$PF_2$: <(account issue), (accounts ledger), (optimised payment strategy)>,
corresponds to Financial Accounting (FI) module,
whereas,
$PF_3$: <(accounts ledger), (process consumption accounting), (corporate budgeting)>,
$PF_4$: <(process consumption accounting), (product cost accounting), (controlling strategy)>,
$PF_5$: <(product cost accounting), (profitability analysis), (central planning strategy)>,
corresponds to Controlling (CO) module.

Accounting deals with the value-based representation of business processes and is tasked with planning, controlling, and checking the value flow in an enterprise. In accordance with the target group, accounting is divided into internal and external accounting.

For example the activities included in the fragment $PF_5$: <(product cost accounting), (profitability analysis), (central planning strategy)>, provide quantitative information to decision makers within the enterprise, and are considered part of the internal accounting process. Moreover the activities included in the fragment $PF_1$: <(order entry), (account issue), (communication strategy)>, refer to an external group of the enterprise and are considered part of the external accounting process.

## 4.1.4 Plants (Netherlands and UK)

### A. Accounting

The accounting implementation at the plants is similar with this at the headquarters we've seen previously.

B. Logistics
The second targeted process is the material logistics process. The enterprise implements mainly the Production Planning (PP) and Material Management (MM) modules, in both plants. The problems it faces are illustrated in the representation below:

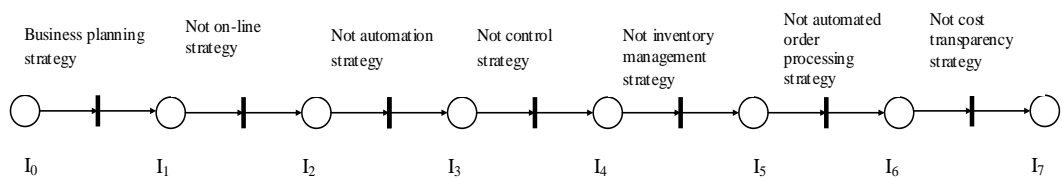

Figure 17: Current state for the logistics process

$I_0$: Purchase requisition, $I_1$: Sales and operations planning, $I_2$: Material requirements planning, $I_3$: Material procurement, $I_4$: Goods receipt, $I_5$: Goods stock, $I_6$: Invoice verification, $I_7$: Pricing.

The problems the enterprise faces are mainly lack of cost transparency within the logistics chain. In the above representation a make-to-stock planning strategy has been applied.

The future state with SAP (To-Be) will be as follows:

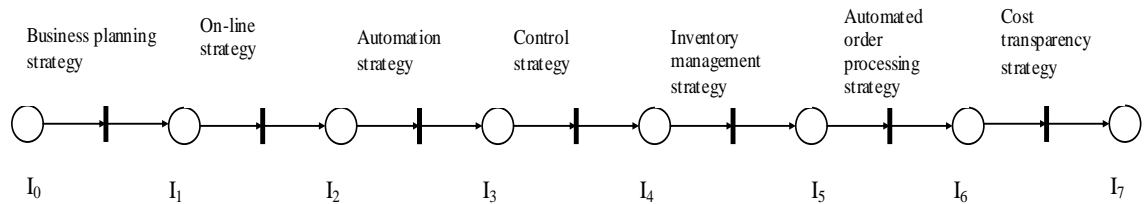

Figure 18: Future state for the logistics process

$I_0$: Purchase requisition, $I_1$: Sales and operations planning, $I_2$: Material requirements planning, $I_3$: Material procurement, $I_4$: Goods receipt, $I_5$: Goods stock, $I_6$: Invoice verification, $I_7$: Pricing.

The logistics components of the R/3 system can be used for different types of production, including make-to-order production, repetitive manufacturing, manufacturing of products with variants, and batch-oriented process manufacturing. The goal is to minimize the costs of all factors that do not contribute directly to the value added process (for example, reducing storage times, warehouse stock, setup times) [Buck-Emden 2000].

We can describe the functionality of the above logistical process: Data from Sales and Distribution, such as analysis of customer needs, can flow directly into the planning operations of the R/3 system's MRP II planning functions. This process chain consists of business planning which generates data for sales planning and the production planning program is derived who serves as the basis for the materials requirement planning and material procurement. The material procurement includes activities such as management of outline agreements and scheduling agreements. In warehousing, after the goods receipt, inventory must be managed on a value basis. Finally the process ends with invoice verification and updating of the value of the material. At this level there is tight link to the Financial Accounting (FI).

From the logistics process presented above, the process fragments: $PF_1$: <(purchase requisition), (sales operations planning), (business planning strategy)>, $PF_2$: <(sales operations planning), (material requirements planning), (on-line strategy)>, $PF_3$: <material requirements planning), (material procurement), (automation strategy)> corresponds to Production Planning (PP) module, whereas, the process fragments: $PF_4$: <(material procurement), (goods receipt), (control strategy)>, $PF_5$: <(goods receipt), (goods stock), (inventory management strategy)> $PF_6$: <(goods stock), (invoice verification), (automated order processing)>, $PF_7$: <(invoice verification), (pricing), (cost transparency strategy)> corresponds to Material Management (MM) ,module.

## 4.1.5 Refinement

In all the above representations we can proceed further into a lower level through refinement. For example in the last illustration of the logistics process for the plants the fragment $PF_1$: <(purchase requisition), (sales operations planning), (business planning strategy)> can be refined into: $PF_{1.1}$: <(purchase requisition), (sales operations planning), (forecasting strategy)>, $PF_{1.2}$: <(purchase requisition), (sales operations planning), (planning for finished products)>. The fragment $PF_{1.1}$ can be further refined into: $PF_{1.1.1}$ <(purchase requisition), (sales operations planning), (mid-term planning strategy)>, $PF_{1.1.2}$ <(purchase requisition), (sales operations planning), (long-term planning strategy)>. The above can be graphically represented as:

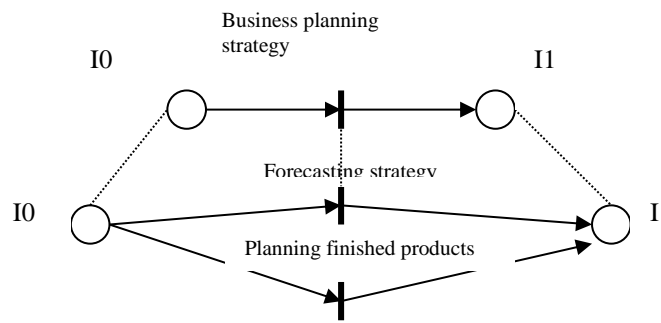

Figure 19: $PF_1$ refinement

## 5. Conclusion and Further Work

In this paper we have seen so far the ontology of the Reusable Organisational Change framework. The modeling of the current state of the enterprise and the proposed SAP solution is in Petri-nets, giving us the opportunity to represent the planning strategy from one state to another.

The levels of abstraction in each phase are three namely intentional, strategical, and operational. We think in terms of goals because organisations think in terms of their objectives [Rolland and Prakash 2000] and not in terms of their processes and functions. Supplementary to this we take into consideration the correspondent strategy in order to align the business processes of SAP to these of the enterprise.

SAP goals are supportive to enterprise goals, and are used to implement enterprise high level objectives such as increase of the profit, which is an objective of every commercial organisation, as well as, improvement of the customer satisfaction, improvement of the quality of service and as a result to this, consistent growth and increase of the market shares. Perhaps every commercial organisation has to choose within a variety of computer systems solutions but when it chooses ERP and more specifically SAP, it must have the possibility of gaining the most from this choice. .

ERP systems are often large systems, of many interacting components. Each component interacts with other components within the system. Thus despite the diversity of the systems we want to model, several common points stand out. These points are modularity and concurrency. Modularity because systems consisting of separate interacting components. Each component may itself be a system and its behaviour can be described independently from other components, apart from well-defined interactions. Concurrency because activities of one component of a system may occur simultaneously with other activities of other components.

We have discussed so far the application of the proposed ROC (Reusable Organisational Change) framework From these case study some conclusions came out. Firstly the way of approach in a real project. SAP is not being developed as a unity, but we have a targeted process. An adjustment in a targeted process is being done and all the necessary components are implemented. So the intention is not to develop SAP itself, but to adjust all its necessary components to facilitate the implementation of a targeted process.

In this project we don't have implementation in a single plant but in the headquarters as well as in two differently located plants. In this case study we can not select components but whole modules because otherwise the project would be chaos. This happens because we need common procedures within the company and SAP offers this commonality.

The above conclusions can be represented in a metamodel, the SAP implementation method metamodel.

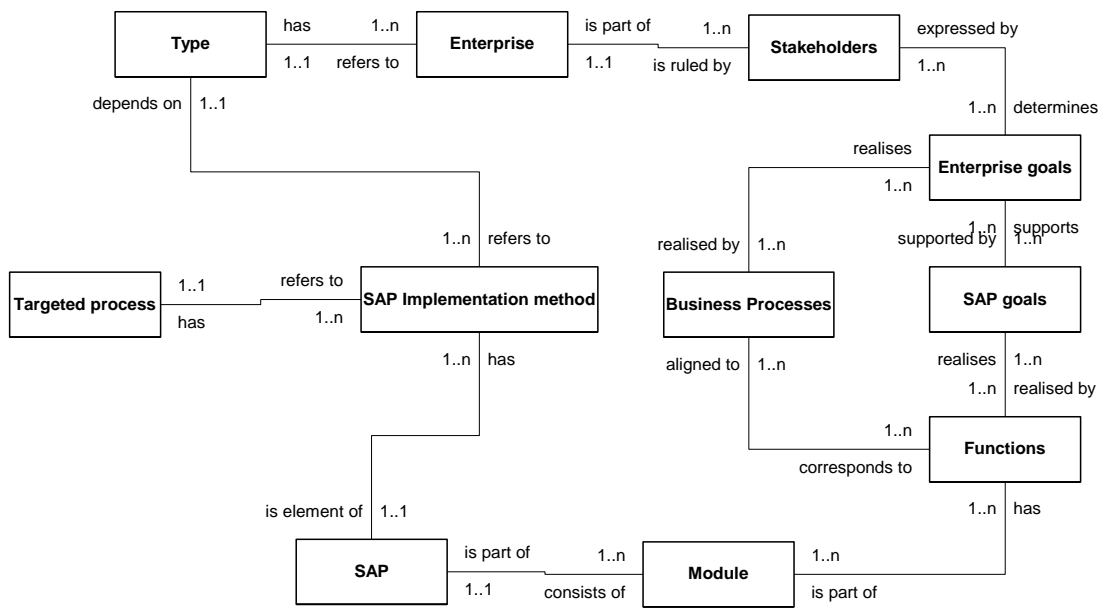

Figure 20: SAP Implementation method metamodel

# 6. References

**Aamodt, A., Plaza, E.** Case-based reasoning: Foundations issues, methodological variations and system approaches. *AI-Communications*, 7(1), 1994, pp 39-59.

**Battacherjee, J.** SAP R/3 Implementation at Geneva Inc. JCAIS, Communications of the Association of Information Systems. Vol. 4 (3), 2000.

**Blain et al.** Using SAP R/3. Que, 1998.

**Buck-Emden, R.** The SAP R/3 System, An Introduction to ERP and Business Software Technology. Addison-Wesley, 2000.

**Hiquet, B**. *SAP R/3 Implementation Guide*. Macmillan Technical Publishing, 1998.

**Keller, G., Teufel, T**., *SAP R/3 Process Oriented Implementation.* Addison-Wesley, 1998.

**Kolezakis, M., Loucopoulos, P**. *Alignment of the SAP requirements to Enterprise requirements?. EMMSAD '03, CAiSE 2003, Velden Austria, 16-17 June.*

**Loucopoulos, P., Karakostas, V**. *Systems Requirements Engineering*. McGraw-Hill, 1995.

**Loucopoulos, P., Kavakli, V.** Enterprise Knowledge Management and Conceptual Modelling, ER 97, 1997

**Peterson, J.L** Petri Net *Theory and the modelling of Systems*. Prentice-Hall, 1981.

**Petri, C**. *Communication with Automata*. New York:Griffiss Air Force Base. Tech. Rep. RADC-Tr-65-377, vol.1, Suppl. 1.

**Ramesh, B., Jarke, M.,** Towards Reference models for Requirements Traceability. *TSE* 27(1) 58-93, 2001.

**Rolland, C., Prakash**, N. Bridging The Gap Between Organisational Needs And ERP Functionality. *Requirements Engineering* 5(3), 2000, pp180-193.

**Scheer, A-W**. *Business process engineering: reference models for industrial enterprises*. Berlin et al., 1994.

**Weihrauch, K.** SAPinfo-Continous Business Engineering, Waldorf, 1996.

**Welti, N.** Successful SAP R/3 Implementation, Practical Management for ERP Projects. Addison-Wesley, 1999.